\documentclass[3p,11pt,authoryear]{elsarticle}  
\journal{Smart Materials and Structures}

\usepackage[pdfborder={0 0 0}]{hyperref}
\usepackage{helvet} 
\usepackage[usenames,dvipsnames]{color}
\usepackage[mathscr]{euscript}
\usepackage{amsmath,amsfonts,latexsym,amssymb}
\usepackage{tikz}
\usepackage{multirow}
\usepackage{graphics,fancyhdr,graphicx}
\usepackage{epstopdf}
\usepackage{soul}
\usepackage[labelfont=bf]{caption}
\usepackage{setspace}
\usepackage{mathtools}
\usepackage{subcaption}
\usepackage[utf8]{inputenc}
\usepackage{nomencl}
\makenomenclature
\usetikzlibrary{trees}
\usepackage{cellspace} 
\usepackage{makecell} 
\setcellgapes{.75pt}
\usepackage{booktabs}

\makeatletter
\let\@oddfoot\@empty
\let\@evenfoot\@empty
\makeatother
\usepackage{amssymb}
\journal{Smart Materials and Structures}
\makeatletter
\let\@oddfoot\@empty
\let\@evenfoot\@empty
\makeatother
\newlength{\figwidth}
\newlength{\subfigwidth}
\newlength{\subfigheight}
\renewcommand{\equationautorefname}{Eq.}

\def\equationautorefname~#1\null{%
	Eq.~(#1)\null
}

\usepackage{multirow}
\usepackage{lineno}
\begin{document}
	\abovedisplayskip=6.0pt
	\belowdisplayskip=6.0pt
	\title{{\bf \large Digitally Controlled Mechatronic Metamaterials for Actively Induced Targeted Bandgaps }}
	
	\author[IITK-ME]{Vivek Gupta\corref{cor1}}
	\ead{v.gupta-2@tudelft.nl}
    \ead{viv02feb1990@gmail.com}
    \author[IITK-ME]{Aditya Natu}
    \ead{A.M.Natu@tudelft.nl}
    \author[IITK-ME]{S. Hassan HosseinNia}
	\ead{S.H.HosseinNiaKani@tudelft.nl}
    \cortext[cor1]{Corresponding author}   
        \address[IITK-ME]{Delft University of Technology, Delft, The Netherlands}

\begin{abstract}

This paper presents an experimental framework for inducing and tuning vibration bandgaps in digitally controlled mechatronic metamaterials. A slender-beam structure instrumented with collocated piezoelectric sensor–actuator pairs distributed periodically along the length is used as the host medium, with decentralized second-order low-pass resonant filter with negative position feedback controllers implemented in real time on an FPGA platform. Unlike conventional approaches that assess bandgap formation through tip displacement, this study relies on bending strain minimization of piezoelectric sensors as the principal indicator of control-induced bandgaps. This reflects more accurately the moment-based phase cancellation dynamics similar to resonator dynamics. We derive analytical expressions for transmissibility in an $n \times n$ decentralized feedback architecture and verify them experimentally using a $7\times 7$ unit-cell configuration. The findings show that resonant controllers with negative feedback applied at the unit-cell level can be systematically tuned through controller gain and damping to open targeted low-frequency bandgaps and significantly improve vibration attenuation. By shifting the focus to localized dynamics, this work deepens the understanding of how control-induced bandgaps emerge and demonstrates a scalable pathway for designing programmable mechatronic metamaterials based on unconventional resonator behavior.
\end{abstract}

\maketitle	

\newpage

\section{Introduction}
\label{section:intro}

The engineering of elastic and acoustic metamaterials has seen major advances in recent years, with increased emphasis on actively tunable systems that can adapt to changing environments or performance criteria, e.g., \cite{liu2000locally,hussein2014dynamics,Sugino2017AMetastructures,failla2024current}. Such designed metamaterials with locally resonant units provide the basis for bandgap formation, enabling the suppression of vibrations and elastic waves within specific frequency ranges \cite{sugino2016mechanism,gupta2020exploring,gupta2022dispersion}. However, conventional passive implementations face limitations in real-time tunability, narrow bandwidth, and non-adaptive behavior, particularly in scenarios demanding dynamic control. To overcome these constraints, research has increasingly focused on digitally controlled mechatronic metamaterials that make use of embedded sensors, actuators (e.g., piezoelectrics), and feedback algorithms for active wave manipulation.

A notable trajectory can be seen in the evolution of integrating piezoelectric transducers with digital feedback control to form programmable unit cells capable of altering their local stiffness and damping characteristics in real time, as demonstrated in existing scientific works. For example, \cite{fang2022valley} and \cite{yuan2019tuning} developed programmable acoustic lattices by synchronizing piezoelectric resonators with digital controllers to create wide and shiftable bandgaps. \cite{lin2023electroelastic} explored tunable phononic crystals with embedded piezo-patches for real-time vibration filtering through metasurfaces with resonant piezoelectric shunts. Similarly, Chen et al.\ (2021) employed active impedance modulation strategies to induce broadband attenuation in elastic beams. \cite{nima2010adaptive} employed second-order low-pass resonant filters with negative position feedback and adaptive gain strategies to precisely control the onset and width of stopbands.

Across many of existing works, the analytical modeling of piezoelectric beam structures has advanced our understanding of tunable bandgaps for vibration control. \cite{huang2009wave} derived dispersion relations in rods with shunted piezo patches, while \cite{hussein2014dynamics} formalized phononic wave modeling in periodic media. \cite{das2025wave} provided exact solutions for sandwich beams with periodic piezo elements. \cite{kaczmarek2024active, kaczmarek2024bandgap} integrated active control into beam bandgap analysis, and \cite{zhu2014negative} offered an effective medium framework for subwavelength wave behavior. However, experimental engineering of bandgap dynamics based on these analytical models remains largely unexplored, which this work aims to address.

Several studies explored the design changes of the host  structure. For example, \cite{wen2020enhanced} shown the periodic variable cross-section design for the acoustic metamaterial beam is an effective measure to obtain more and broader bandgaps. Using the coupling mechanisms of the Bragg scattering and locally resonant bandgaps and a metamaterial beam with periodically variable cross-sections is designed to enhance the vibration reduction capacity in wide frequency ranges. 
\cite{wu2022design} developed a metamaterial beam with high specific energy absorption coefficient using variable cross section. Their model optimized equal-section beam to the variable section using chained beam constraint model (CBCM) and the improvement in specific energy absorption is observed with variable section of the beam.
\cite{sugino2016mechanism} investigated a metastructure consisting of a beam substrate with a periodic arrangement of spring-mass resonators. They showed the formation of a bandgap centered around the resonant frequency of the system, with the bandgap width being influenced by the ratio of the resonator mass to the mass of the beam segment corresponding to each unit cell. Similarly, \cite{huang2016attenuation} examined a beam incorporating more complex resonators, consisting of inclined trusses integrated with conventional spring-mass resonators. Despite the increased complexity, the bandgaps were still observed near the natural frequencies of the spring-mass system.

While recent studies, such as \cite{kaczmarek2024creating}, have made important progress in demonstrating digitally controlled elastic metamaterials, they stop short of unpacking how bandgaps actually emerge at the scale of individual units, or how resonant control dynamics at that level translate into system-wide attenuation. In this work, we take a step closer to that fundamental question by experimentally demonstrating, at the unit-cell scale, how resonant feedback can actively induce and tune targeted bandgaps. A key distinction of our approach is the choice of metric: instead of relying on transverse tip displacement, as is commonly done, we examine the minimization of bending strain in the piezoelectric actuators. This perspective more directly reflects the phase cancellation caused by moment-based actuation and provides a clearer window into the physical origin of control-induced bandgaps. By reframing how these effects are observed and validated, our study uncovers new dynamics at play and provides a pathway toward practical, fine-grained control of programmable mechatronic metamaterials.

The remainder of this paper is organized as follows. Section~\ref{sec:2} presents the experimental modeling framework, beginning with the architecture of the mechatronic metamaterials, details of the structure and hardware, and the controller implementation. This is followed by an analytical framework for the $n \times n$ feedback design, including derivations of the transmissibility ratio and an experimental case study on a $7 \times 7$ unit cell. Section~\ref{sec:3} reports the experimental results and the observed bandgaps, with emphasis on the influence of controller gains and damping. Section~\ref{sec:4} provides a broader discussion of the findings, and Section~\ref{sec:5} concludes the paper with final remarks and outlook.

\section{Experimental Modeling of Bandgaps through mechatronic approach}\label{sec:2}
This section introduces the structure and simplified model of the active piezoelectric metamaterial beam, followed by an overview of the control architecture and implemented controllers. Finally, algebraic expressions are derived to demonstrate the targeted and enhanced bandgaps achieved through different control strategies. Schematic representation of a single unit cell integrated with a collocated actuator–sensor with the representation of a metabeam with the $j^{\text{th}}$ unit cells distributed along its length is shown in \autoref{fig: schematic1}.

\subsection{Structure and Hardware}

The experimental setup of the studied structure consists of a slender aluminum alloy beam, assumed in simulations with a density of 
$\rho_s = 2700 \,\text{kg/m}^3$ and an elastic modulus of $c_s = 69 \,\text{GPa}$. 
The beam is rigidly clamped at its base using the fixtures, as shown in \autoref{fig:Exp setup}. The vibrational response at both the tip and the clamped base of the beam is captured using a pair of piezoelectric transducers. These signals are subsequently processed to determine the system transmissibility and evaluate its performance.

Collocated piezoelectric patch actuators and sensors (PI P-876.SP1 DuraAct) are mounted on the beam, with seven sensor--actuator pairs implemented to generate the desired bandgaps. The piezo patches serve in both actuation and sensing roles. Signal conditioning is carried out using custom-built charge amplifiers based on TL074 operational amplifiers. In this implementation, the feedback network is characterized by $C_f = 200 \,\text{nF}$ and $R_f = 2 \,\text{M}\Omega$. The resulting transfer function between the piezoelectric sensor charge and the amplifier voltage output can be expressed as

\begin{equation}
    \frac{V_0}{Q} = \frac{-R_f s}{\left(R_f C_f s + 1\right)\left(R_i(C_p + C_s)s + 1\right)}.
    \label{eq:transfer_function}
\end{equation}
The transfer function has two poles at $\omega_{1} = 1/(R_f C_f)$ and $\omega_{2} = 1/[R_i(C_c + C_p)]$, with a flat-band gain of $1/C_f$. Each actuator is driven by a Dual-Channel 300V Amplifier (BD300). The controllers are digitally implemented on an NI cRIO-9039 FPGA with a 10~kHz sampling frequency, which also monitors and records performance signals. Data acquisition modules include NI9215 for charge amplifier outputs, NI9234 for acceleration measurements, and NI9264 for generating excitation signals for the shaker and piezoelectric actuators.

\subsection{Mechatronic Metamaterials Architecture}
The complete metamaterial architecture is generated once the beam is covered with \(N\) repeated collocated piezoelectric transducers along its length. In this configuration, all the piezoelectric patches are poled in the thickness direction. A visual representation of such a structure with $n$ unit cells is shown in Fig.~1. Here, for $i = 1,\ldots,n$ defines the unit cell considered, \(V_{si}\) is the voltage produced by the \(i\)-th sensor, \(V_{ai}\) is the voltage sent to the \(i\)-th actuator, and \(C_i(s)\) denotes the identical controller implemented for each unit cell \(i\). Specifically, the patch sensor output is related to the average beam curvature at its location, whereas the actuator produces a pair of bending moments with amplitudes proportional to the applied voltage.

\begin{figure}[ht!]
  \centering
  \includegraphics[width=\textwidth]{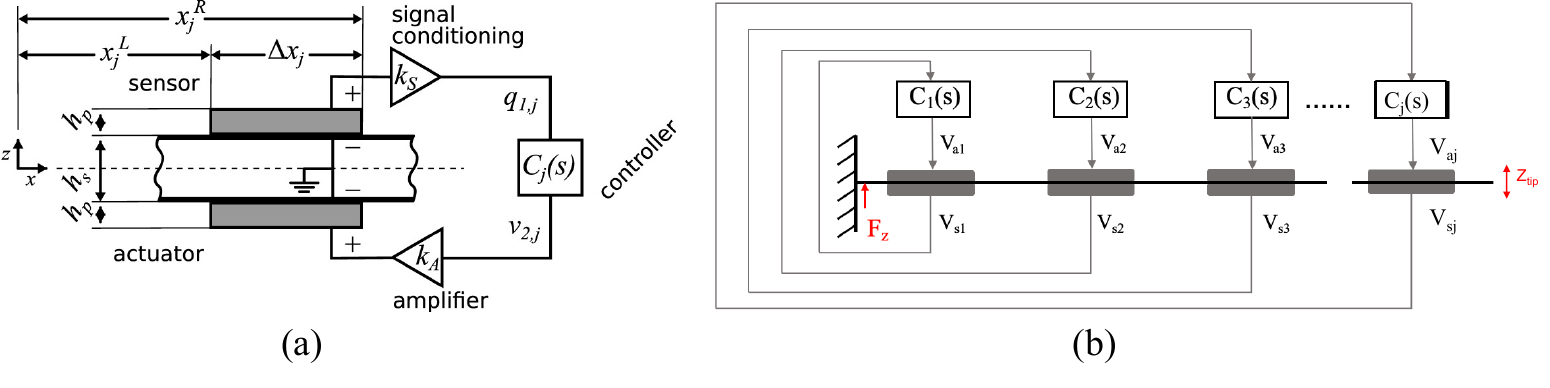}
\caption{(a) Schematic representation of a single unit cell integrated with a collocated actuator–sensor configuration driven through the controller, and (b) representation of a metabeam with the $j^{\text{th}}$ unit cells distributed along its length \cite{kaczmarek2024bandgap}.}

  \label{fig: schematic1}
\end{figure}
The transfer function from actuator 1 to sensor 7 is of primary interest for analyze system transmissibility for vibrational bandgaps measurement. This study is critical for assessing attenuation frequency region through transmission, and, consequently, bandgap performance. The dynamics from actuators-sensors, and their cross-coupling therefore define the performance channel, within which bandgaps are targeted.  The collocated channel is defined by the transfer function from actuator voltage $V_{ai}$ to sensor voltage $V_{si}$. Controllers $C_i(s)$ are applied to these collocated channels with the objective of inducing bandgaps in the performance channel.  The considered configuration of mechatronic metamaterial is shown Fig.~\ref{fig: schematic1}.

\begin{figure}[ht!]
  \centering
  \includegraphics[width=\textwidth]{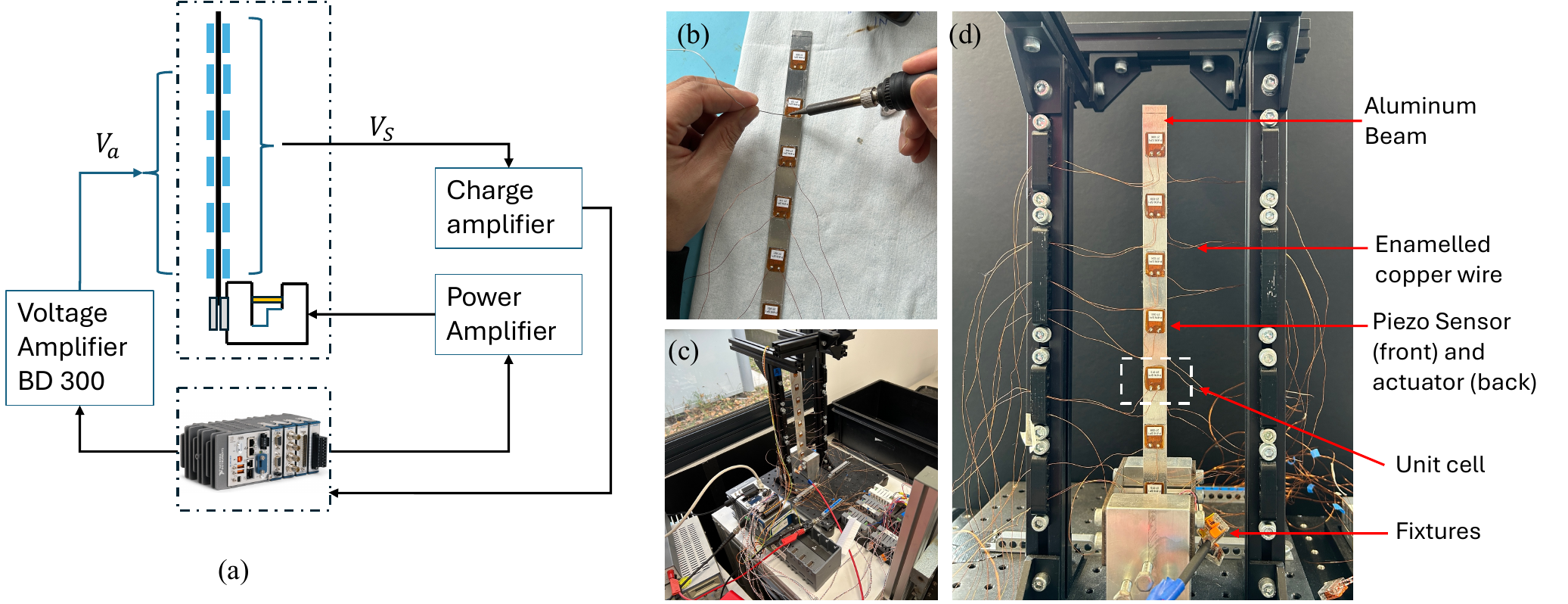}
  \caption{(a) Fabrication of mechatronic metamaterial beam with piezo actuators and sensors attached back to back surface of the beam, performing soldering, (b) A complete experimental setup for measurement with controllers and testbed, (c) Close-up of beam.}
\label{fig:Exp setup}
\end{figure}
\subsection{Controller Implementation}
\begin{figure}[ht!]
  \centering
  \includegraphics[width=0.65\textwidth]{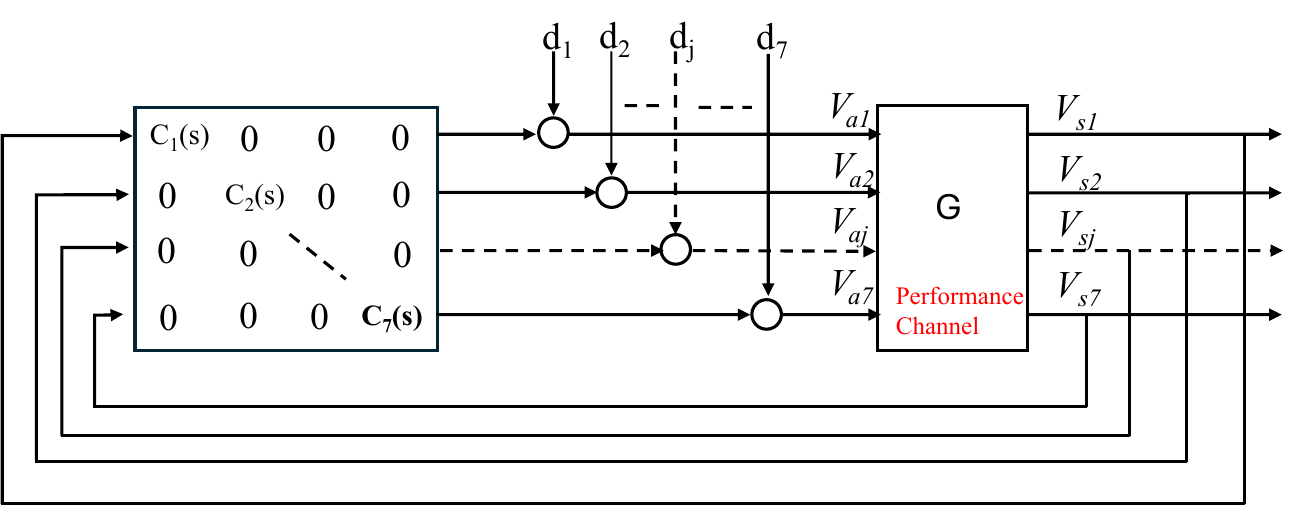}
  \caption{Closed-loop block diagram of the decentralised control architecture
used in active piezoelectric metamaterials.}
\label{fig:block_diagram}
\end{figure}

Active control for generating locally resonant bandgaps in beamlike piezoelectric metastructures requires: (1) a sensor to measure vibrations, (2) an actuator to suppress them, and (3) a control system to close the loop. Each unit cell forms an independent control loop acting as a local resonator. A decentralized control strategy is adopted, where each sensor-actuator pair operates independently, enhancing robustness and enabling parallel computation. Figure~\ref{fig:block_diagram} illustrates the closed-loop block diagram. Controllers \( C_i(s) \) aim to minimize the influence of base excitation \( F_z \) on tip displacement \( z_{\text{tip}} \). The plant \( G \) represents the metastructure, and \( d_i \) denotes disturbances within each unit cell. The controller matrix is block-diagonal, reflecting the absence of inter-cell coupling.
Given such control architecture, the controllers \( C_i(s) \) are pivotal in defining the unit cell dynamics and generating a bandgap. 



\subsection{Analytical Framework for \texorpdfstring{$n \times n$}{n x n} Feedback Design}

To quantify the effect of local feedback on system behavior, we consider a general multivariable plant \( G(s) \in \mathbb{R}^{n \times n} \), where each element \( G_{ij}(s) \) represents the transfer function from actuator \( j \) to sensor \( i \). The control architecture is decentralized, with a diagonal controller matrix:
\begin{equation}
C(s) = \mathrm{diag}(C_1(s), C_2(s), \dots, C_n(s))
\end{equation}
Each controller \( C_i(s) \) is implemented as a second-order low-pass filter:
\begin{equation}
C_i(s) = \frac{g_i}{\dfrac{s^2}{\omega_c^2} + 2 \zeta_{c_i} \dfrac{s}{\omega_c} + 1}
\end{equation}
where:
\begin{itemize}
    \item \( \omega_c \) is the fixed resonant frequency where the bandgap is targeted,
    \item \( g_i \) and \( \zeta_{c_i} \) are controller gain and damping, tuned individually per unit,
    \item The gains satisfy the condition \( g_i < G_{ii}(0)^{-1} \) to ensure stability and passivity.
\end{itemize}

Each controller is tuned based solely on its corresponding local plant \( G_{ii}(s) \). This allows modular tuning while accounting for minor variations due to non-ideal collocation or dynamics.

The closed-loop response of the system under external actuation is described by:
\begin{equation}
X(s) = \left(I + C(s)G(s)\right)^{-1} G(s) d(s)
\end{equation}
where \( d(s) \in \mathbb{R}^n \) is the vector of disturbance inputs and \( X(s) \in \mathbb{R}^n \) is the vector of sensor outputs.

To evaluate vibration transmission, we apply excitation only at actuator 1 to emulate the base excitation scenario:
\begin{equation}
d(s) = [d_1(s), 0, \dots, 0]^T
\end{equation}
yielding:
\begin{equation}
X(s) = d_1(s) \cdot \left(I + C(s)G(s)\right)^{-1} G_{:,1}(s)
\end{equation}

\subsubsection{Analytical expression for the transmissibility ratio}
We partition \( G(s) \) as:
\begin{equation}
G(s) =
\begin{bmatrix}
G_{11}(s) & g_{12}^\top(s) \\
g_{21}(s) & G_{22}(s)
\end{bmatrix}
\end{equation}
where \( g_{12}(s), g_{21}(s) \in \mathbb{R}^{n-1} \) and \( G_{22}(s) \in \mathbb{R}^{(n-1)\times(n-1)} \).

Since all unit cells in the metamaterial beam are physically identical and instrumented with collocated actuator–sensor pairs of the same type, their local transfer functions \( G_{ii}(s) \) are nominally the same. Minor variations caused by manufacturing tolerances or sensor placement are negligible compared to the dominant dynamic characteristics. It is therefore reasonable to assume that all controllers share identical parameters, i.e.
\begin{equation}
C_i(s) = C(s), \quad \forall i=1,\dots,n.
\end{equation}
This assumption simplifies the analysis while still accurately capturing the feedback behavior of the structure. It also reflects the practical implementation, where a single controller design is duplicated across all unit cells for ease of tuning and hardware uniformity.

For this identical-controller case, define
\begin{equation}
\eta(s) = \left(I_{n-1} + C(s) G_{22}(s)\right)^{-1} g_{21}(s),
\end{equation}
and let \( e_{n-1} \) be the unit vector in \( \mathbb{R}^{n-1} \) selecting the last entry (corresponding to sensor \( n \)). The closed-form transmissibility ratio is
\begin{equation}
\frac{x_n(s)}{x_1(s)} =
\frac{e_{n-1}^\top \, \eta(s)}
{\,G_{11}(s) - C(s) \, g_{12}^\top(s) \, \eta(s)\,}.
\label{eq:trans_ratio}
\end{equation}
This result follows from block elimination of the closed-loop equations and avoids forming the full inverse \( (I + C(s) G(s))^{-1} \).

\subsubsection{Experimental Case: \texorpdfstring{$7 \times 7$}{n x n} Unit Cell}\label{section: controller rule}

In the experimental setup, a flexible aluminum beam is instrumented with \( n = 7 \) collocated piezoelectric actuator-sensor pairs. The plant transfer matrix \( G(s) \in \mathbb{R}^{7 \times 7} \) is identified via experimental frequency response measurements. Each diagonal term \( G_{ii}(s) \) is used to independently tune \( C_i(s) \), implementing the decentralized control law.

Once the controllers are implemented, a chirp signal is injected through actuator 1 while all controllers remain active. The measured output \( X(s) \) is used to compute the transmissibility vector:
\begin{equation}\label{eq: T_s}
T(s) = \left( I + C(s) G(s) \right)^{-1} G_{:,1}(s)
\end{equation}
and each element is normalized by the first component to give:
\begin{equation}
\frac{x_k(s)}{x_1(s)} = \frac{T_k(s)}{T_1(s)}, \quad k = 2, \dots, 7.
\end{equation}

For \( n=7 \), write
\begin{equation}
\eta(s)=\left(I_{6}+C(s)G_{22}(s)\right)^{-1}g_{21}(s),\qquad e_{6}=[0,0,0,0,0,1]^\top,
\end{equation}
so that the end-to-end transmissibility is
\begin{equation}\label{eq: end-to-end}
\frac{x_7(s)}{x_1(s)}=\frac{e_{6}^\top \eta(s)}{\,G_{11}(s)-C(s)\,g_{12}^\top(s)\eta(s)\,}.
\end{equation}

The expressions in \eqref{eq:trans_ratio} (and their $7\times 7$ specialization above) provide analytic predictions for the measured transmissibility, which will be used to show the bandgap experimentally.

\section{Experimental Results and Observed Bandgaps}\label{sec:3}

\begin{figure}[ht!]
  \centering
  \includegraphics[width=\linewidth]{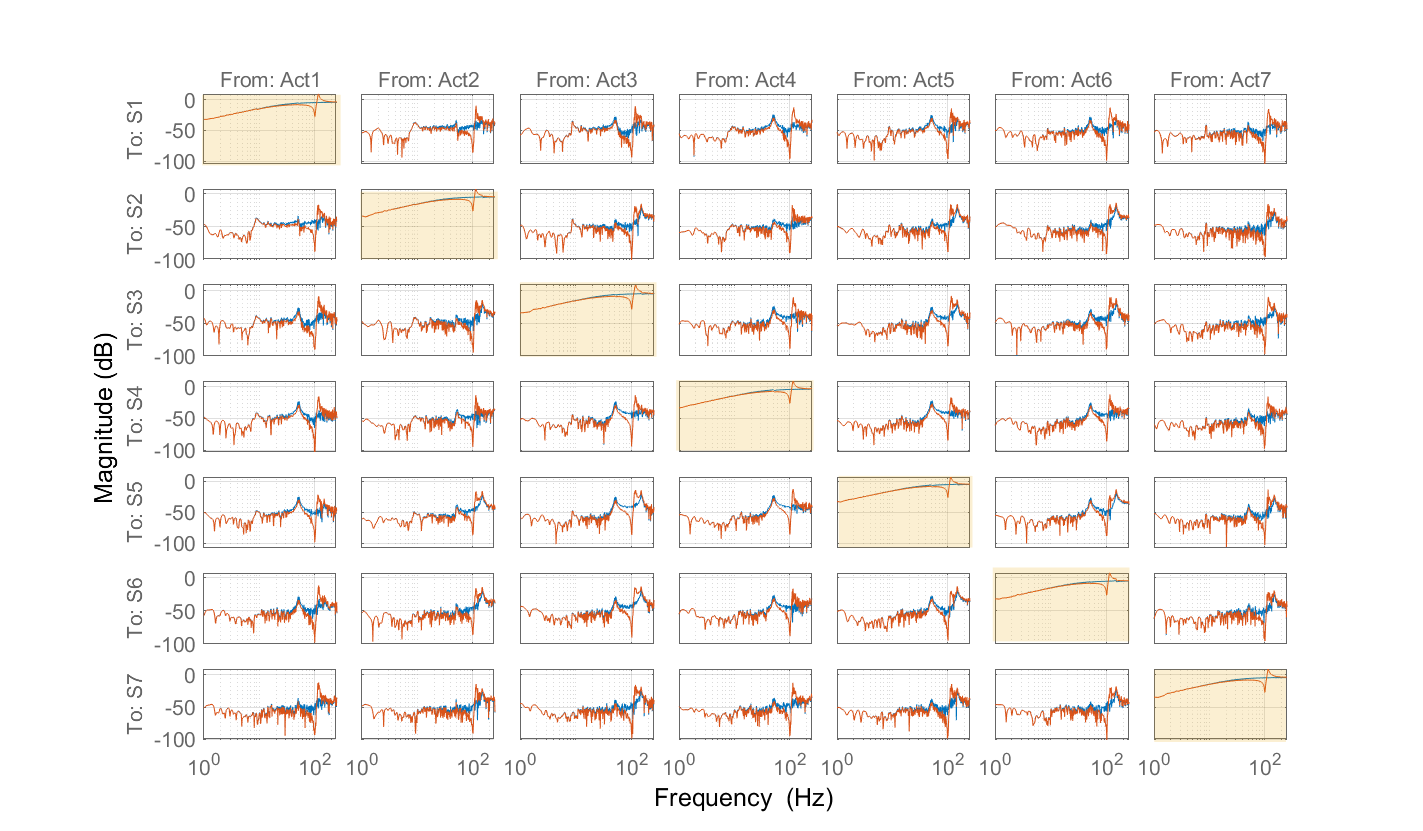}
  \caption{Experimental frequency response magnitudes of the considered $7\times 7$ system, open-loop (blue) and simulated closed-loop responses obtained by applying a resonant controller to collocated pairs (red). The collocated responses are diagonally highlighted in yellow.}
  \label{fig: Open loop}
\end{figure}

The identified MIMO plant is shown in \autoref{fig: Open loop}, where only magnitude responses are presented for clarity. The system is represented by the transfer matrix \(G(s)\), with inputs corresponding to actuator voltages \(V_{a1}, \dots, V_{a7}\) (columns) and outputs corresponding to sensor signals \(V_{s1}, \dots, V_{s7}\) (rows). Thus, each element \(G_{ij}(s)\) denotes the transfer function from actuator \(V_{ai}\) to sensor \(V_{sj}\); for instance, \(G_{11}(s)\) represents the path from \(V_{a1}\) to \(V_{s1}\), \(G_{77}(s)\) from \(F_z\) to \(z_\text{tip}\), and \(G_{36}(s)\) from \(V_{a6}\) to \(V_{s3}\).

The resonant dynamics are applied to each individual collocated unit cell following the rules described in the previous section~\ref{section: controller rule}, from \autoref{eq: T_s} to \autoref{eq: end-to-end}. The corner frequency of the controller is tuned to lie in the vicinity of the desired bandgap region, and the control is implemented in the frequency domain, as illustrated in \autoref{fig: controller_imp}. Here, the controllers act as distributed virtual resonators coupled through the host beam dynamics. Each controller introduces a localized anti-resonance, and their combined action along the beam modifies the global transfer function, transforming isolated notches into a continuous frequency band of reduced transmissibility. This coordinated interaction among controlled unit cells is responsible for the emergence of bandgaps in the closed-loop system. Experimentally designed bandgaps at different frequencies, ranging from high to low, along with the influence of controller dynamics on the attenuation width, are shown in \autoref{fig: bandgap}. The influence of the resonant controller progressively diminishes with decreasing frequency, as the active metabeam system inherently exhibits high-pass characteristics that directly affect bandgap formation, particularly its width. As a result, the attenuation associated with the bandgaps diminishes nonlinearly toward lower frequencies and displays a quasi-exponential decay trend in the response magnitude.

\begin{figure}[ht!]
  \centering
  \includegraphics[width=\linewidth]{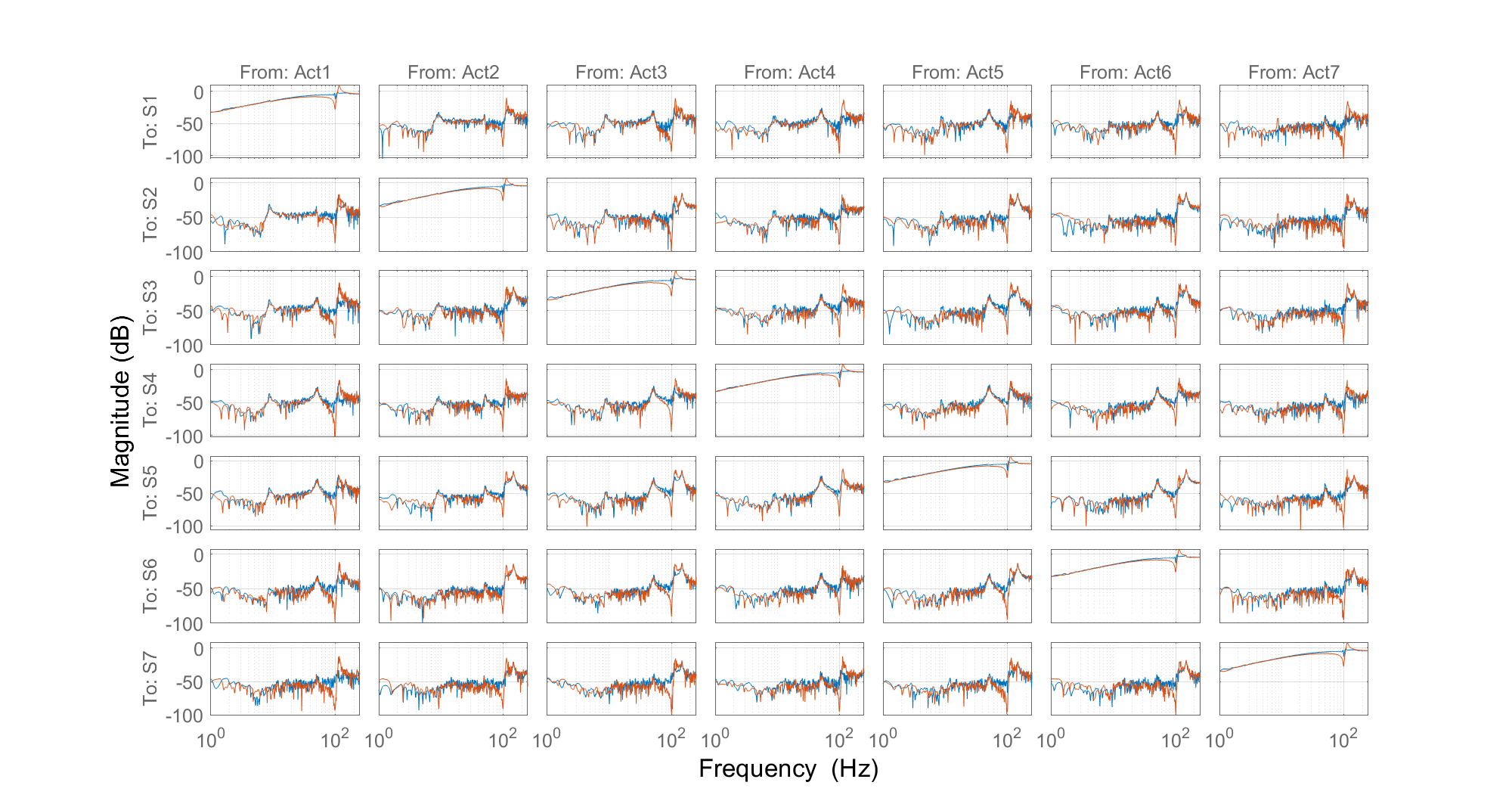}
  \caption{Experimental validation of the closed-loop response (blue) compared with the simulated closed-loop response (red).}
   \label{fig: Closed loop}
\end{figure}

\begin{figure}[ht!]
  \centering
  \includegraphics[width=0.9\linewidth]{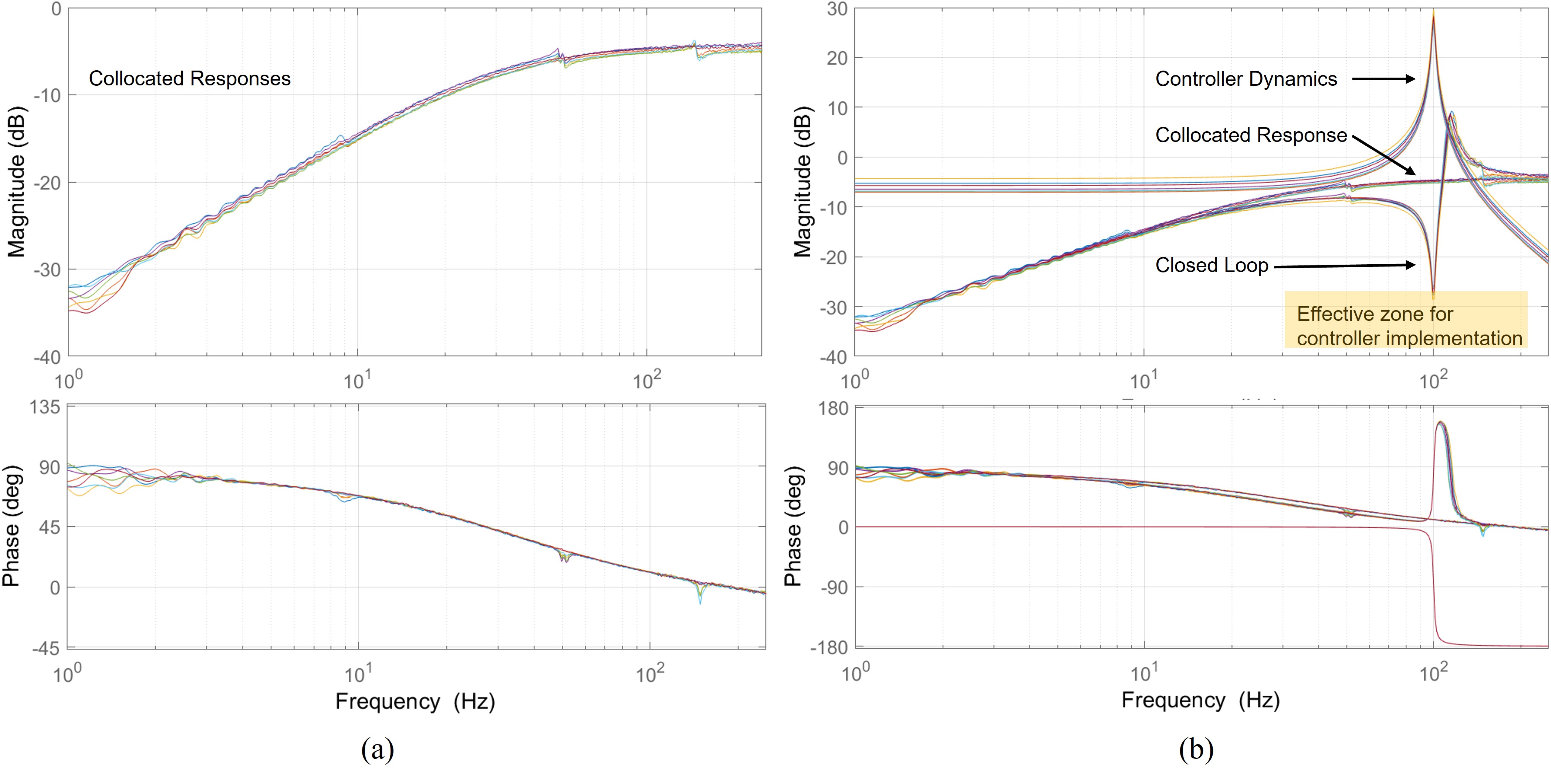}
\caption{Bode magnitude and phase responses of all collocated actuator–sensor transfer functions (a).  
The effect of second-order low-pass resonant filter with negative position feedback controllers implemented on the respective collocated transfer functions is also illustrated (b), demonstrating the achievable vibration attenuation and phase modification.}

  \label{fig: controller_imp}
\end{figure}

\begin{figure}[ht!]
  \centering
  \includegraphics[width=0.95\textwidth]{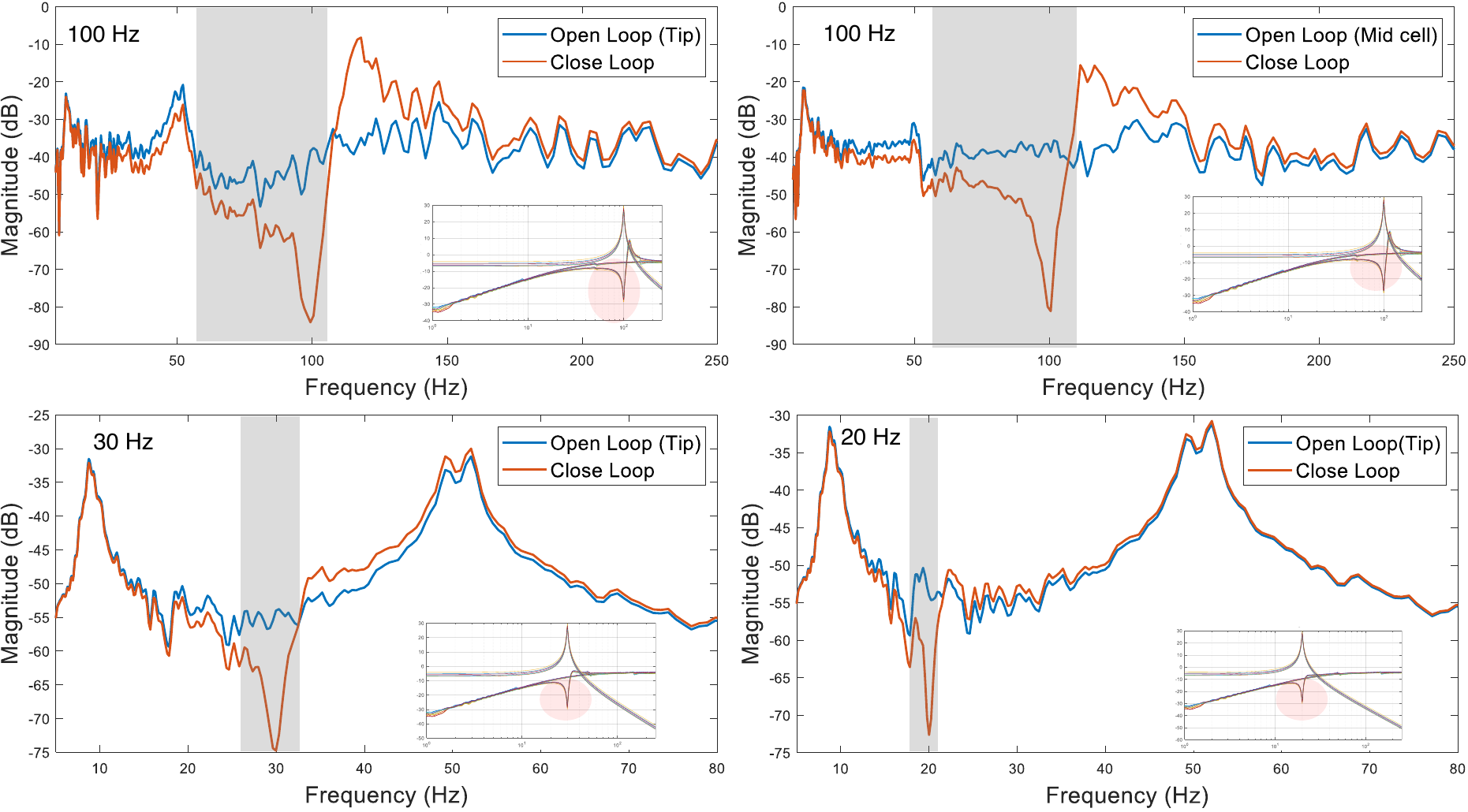}
  \caption{Experimentally designed vibration bandgaps at different frequencies. Responses without controller dynamics are shown in blue, and responses with the controller applied are shown in red.}
  \label{fig: bandgap}
\end{figure}

\subsection{Effect of the controller gains and damping}
\begin{figure}[t!]
  \centering
 \includegraphics[width=0.9\textwidth]{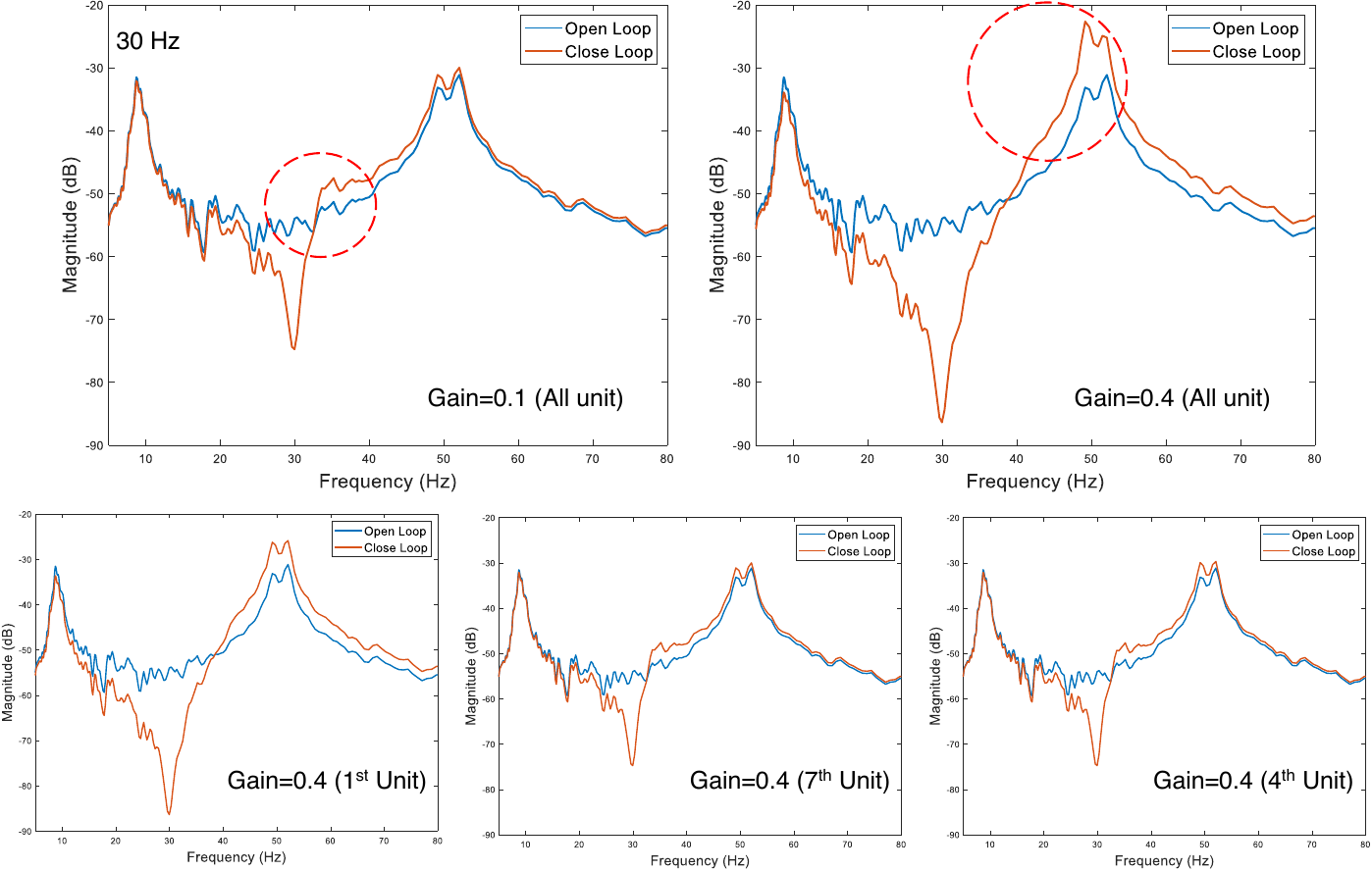}
  \caption{Effect of the controller gain parameter on the low-frequency response of the system.}
\label{fig: gain effect}
\end{figure}

The selection of controller gains in the second-order low-pass resonant filter with negative position feedback plays a crucial role in shaping the bandgap characteristics. High feedback gains increase the virtual stiffness and damping around the targeted resonant modes, leading to sharper and wider attenuation zones in the frequency response. However, excessively high gains result in controller--structure interactions, introducing spillover or modal instability outside the intended frequency range, as shown in \autoref{fig: gain effect}. The gain effect across different unit cells is directly influenced by the proximity of their dominant modal frequencies to the targeted corner frequency or bandgap. For example, when the targeted bandgap lies near the first modal frequency, applying a high control gain is effective primarily in the first unit cell, as shown in \autoref{fig: gain effect}a. Consequently, additional control effort or power input in the remaining cells becomes redundant, as depicted in \autoref{fig: gain effect}b and c. The same reasoning extends to higher frequency ranges, where only those unit cells whose modal frequencies lie close to the targeted bandgap region significantly contribute to the control performance.

Here, the spatial distribution of mode shapes dictates how vibrational energy is localized along the structure. When the targeted frequency lies close to a particular mode, the associated deformation pattern dominates the system response. Unit cells located near the anti-nodes of that mode experience the largest bending strain and, therefore, interact most effectively with the control system. In contrast, unit cells near nodal regions contribute negligibly to the overall response, making additional control gain or transducer activation energetically inefficient. Consequently, for each targeted modal range, the effective number of unit cells (or transducers) corresponds closely to the number of dominant modes participating in the dynamic response, which is consistent with the observations of \cite{kaczmarek2024active} for low-frequency bandgaps in cantilevers.
The damping ratio, embedded in the controller's transfer function, directly influences the smoothness and depth of the attenuation band, affecting both amplitude suppression and transient behavior. Optimal damping must be tuned to suppress resonances effectively while avoiding excessive bandwidth overlap that can blur distinct bandgaps. Experimental observations from the piezo-distributed metabeam reveal that precise calibration of gain--damping pairs leads to localized modal suppression and shifts in peak transmissibility, providing a programmable mechanism for vibration rejection. Thus, controller design is not merely a stabilizing tool but a tunable parameter space that governs the effective dispersion relation and wave attenuation landscape of the smart metamaterial.

\section{Discussion}\label{sec:4}
The transmissibility ratio provides \autoref{eq:trans_ratio} a quantitative measure of how decentralized resonant controller with negative feedback modifies vibration transmission relative to the (uncontrolled) plant structural dynamics. Normalizing the closed-loop transmissibility
\[
T_k^{cl}(s)=\frac{e_{k-1}^T\eta}{G_{11}-C\,g_{12}^T\eta},
\qquad \eta=(I_6+C G_{22})^{-1}g_{21},
\]
by the scalar $e_{k-1}^T\eta$ yields
\[
T_k^{cl}(s)=\frac{1}{\alpha_k(s)-C(s)\beta_k(s)},
\quad
\alpha_k:=\frac{G_{11}}{e_{k-1}^T\eta},\quad
\beta_k:=\frac{g_{12}^T\eta}{e_{k-1}^T\eta}.
\]
This form shows that attenuation is governed by the magnitude of the effective denominator
$D_k(s):=\alpha_k(s)-C(s)\beta_k(s)$, since small $|T_k^{cl}|$ is equivalent to large $|D_k|$.
Inter-cell coupling enters through $G_{22}$ in $\eta=(I_6+C G_{22})^{-1}g_{21}$ and therefore directly
shapes both $\alpha_k$ and $\beta_k$. In particular, stronger coupling modifies $\eta$ such that the
feedback-driven interaction ratio $\beta_k$ is enhanced and/or becomes more favorably phased with the
resonant controller $C(s)$ near the tuning frequency. As a result, $|D_k(j\omega)|$ increases over the
targeted frequency interval, leading to reduced $|T_k^{cl}(j\omega)|$ and stronger bandgap attenuation, as observed experimentally in the Figs. \ref{fig: controller_imp} and \ref{fig: gain effect}.

In summary, the closed-loop transmissibility provides a rigorous metric for evaluating the effectiveness of decentralized resonant feedback in suppressing vibration transmission. Reduced closed-loop transmissibility over targeted frequency intervals directly indicates the formation of active bandgaps and enhanced vibration isolation achieved through local resonant control.
Bandgaps are also influenced by the intrinsic capacitance $C$ of the piezoelectric material. When combined with a resistive element $R$ (e.g., data acquisition hardware), this forms a series $RC$ network that exhibits high-pass (HP) characteristics with corner frequency $\omega_{hp} = \frac{1}{2\pi RC}$.

This HP behaviour leads to parasitic dynamics at low frequencies, directly influencing bandgap formation in active piezoelectric metamaterials. The resonant dynamics of the controller are most effective when the collocated transfer function approaches the 0 dB line, where the characteristic dip appears, as shown \autoref{fig: controller_imp}. The observed dip in the collocated frequency response is therefore consistent with the HP behaviour and can be exploited for bandgap design. At higher frequencies, beyond the dominant piezoelectric dynamics where the HP effect is negligible, but at low frequencies the dip vanishes. To restore and enhance this dip, higher controller gain values are employed, thereby reinforcing the attenuation in the collocated response. The combined action of multiple controllers distributed along the structure then produces a targeted bandgap. Importantly, this approach demonstrates that the controller corner frequency need not coincide with a structural resonance to achieve low-frequency bandgap formation.


\section{Conclusion}\label{sec:5}

The main findings of this study can be summarized as follows:
\begin{itemize}
    \item We experimentally demonstrated digitally controlled mechatronic metamaterials with collocated piezoelectric sensor–actuator pairs, showing that second-order low-pass resonant filters with negative position feedback can successfully generate targeted vibration bandgaps, including in low-frequency regions where passive methods are fundamentally limited.

    \item The effectiveness of the resonant controller was confirmed through systematic tuning of gain, damping, and resonance frequency, enabling programmable and frequency-selective attenuation zones along the metabeam.

\item Analytical and experimental results show strong agreement. The derived closed-loop transmissibility framework provides a rigorous theoretical basis that matches experimental observations and offers a practical tool for shaping and interpreting bandgap behavior by explicitly accounting for inter-cell coupling and localized dynamic effects. This framework further serves as a predictive tool for active bandgap design, as reduced closed-loop transmissibility over targeted frequency intervals directly explains the emergence of active bandgaps and the enhancement of vibration isolation achieved through decentralized resonant control.

    \item Bandgaps were shown to arise even without reliance on structural resonances, driven instead by controller-induced virtual dynamics and parasitic cross-coupling effects. This finding opens new pathways toward adaptive control strategies, multi-dimensional grading, and reconfigurable metamaterial architectures capable of real-time wave manipulation.
\end{itemize}

\section*{Author Contributions}
\noindent
\textbf{Vivek}: Conceptualization; Methodology; Software; Validation; Formal analysis; Investigation; Data curation; Visualization; Writing – original draft; Writing – review \& editing.
\textbf{Aditya}: Investigation; Data curation; Validation.
\textbf{Hassan}: Conceptualization; Methodology; Supervision; Writing – original draft; Writing – review \& editing; Project administration.

\section*{Data Availability Statement}
The data that support the findings of this study are available upon reasonable request from the authors.
\section*{Acknowledgements}
This work was supported by the Netherlands Organization for Scientific Research (NWO) under the HTSM Applied and Technical Science Program, project \textit{MetaMech}, grant number 17976.
\bibliographystyle{elsarticle-harv}
\bibliography{references}


\end{document}